\documentclass[slac_one]{revtex4}
\usepackage{graphicx}
\usepackage{fancyhdr}
\newcommand {\m}          {\rm \, m}
\newcommand {\cm}         {\rm \, cm}
\newcommand {\V}          {\rm \, V}
\newcommand {\ns}         {\rm \, ns} 

\begin{document}

\title{The construction and commissioning of the CMS Silicon Strip Tracker} 

\author{Giacomo Sguazzoni (for the CMS Silicon Strip Tracker Collaboration)}
\affiliation{INFN Sezione di Firenze, via Sansone, 1 - I-50019 Sesto F.no (FI) - ITALY}

\begin{abstract}
As the start up date for LHC approaches, the detectors are readying for
data taking. Here a review will be given on the construction phase
with insights into the various difficulties encountered during the
process. An overview will also be given of the commissioning strategy
and results obtained so far.

The CMS tracker is the largest silicon microstrip detector ever built.
Consisting of three main subsystems, Inner Barrel and Disks, Outer Barrel and End
Caps, it is $5.4\m$ long and is $2.4\m$ in diameter. Total detector surface is
an unprecedented $200\m^2$ with more than 15000 detector modules.

The various integration procedures and quality
checks implemented are briefly reviewed.
Finally an overview is given of checkout procedures
performed at CERN, after the final underground installation
of the detector. 
\end{abstract}

\maketitle

\thispagestyle{fancy}

\section{CONSTRUCTION AND INSTALLATION}

The CMS Silicon Strip Tracker (SST) is the world's largest Silicon
strip detector with a volume of approximately $23\m^3$ instrumented by 15\,148 modules for a
total of 198$\m^2$ of Silicon active area and 9\,316\,352 channels with
full optical analog readout~\cite{cms, TkTDR}.
\begin{figure}[b]
\begin{center}
\includegraphics*[width=0.45\textwidth]{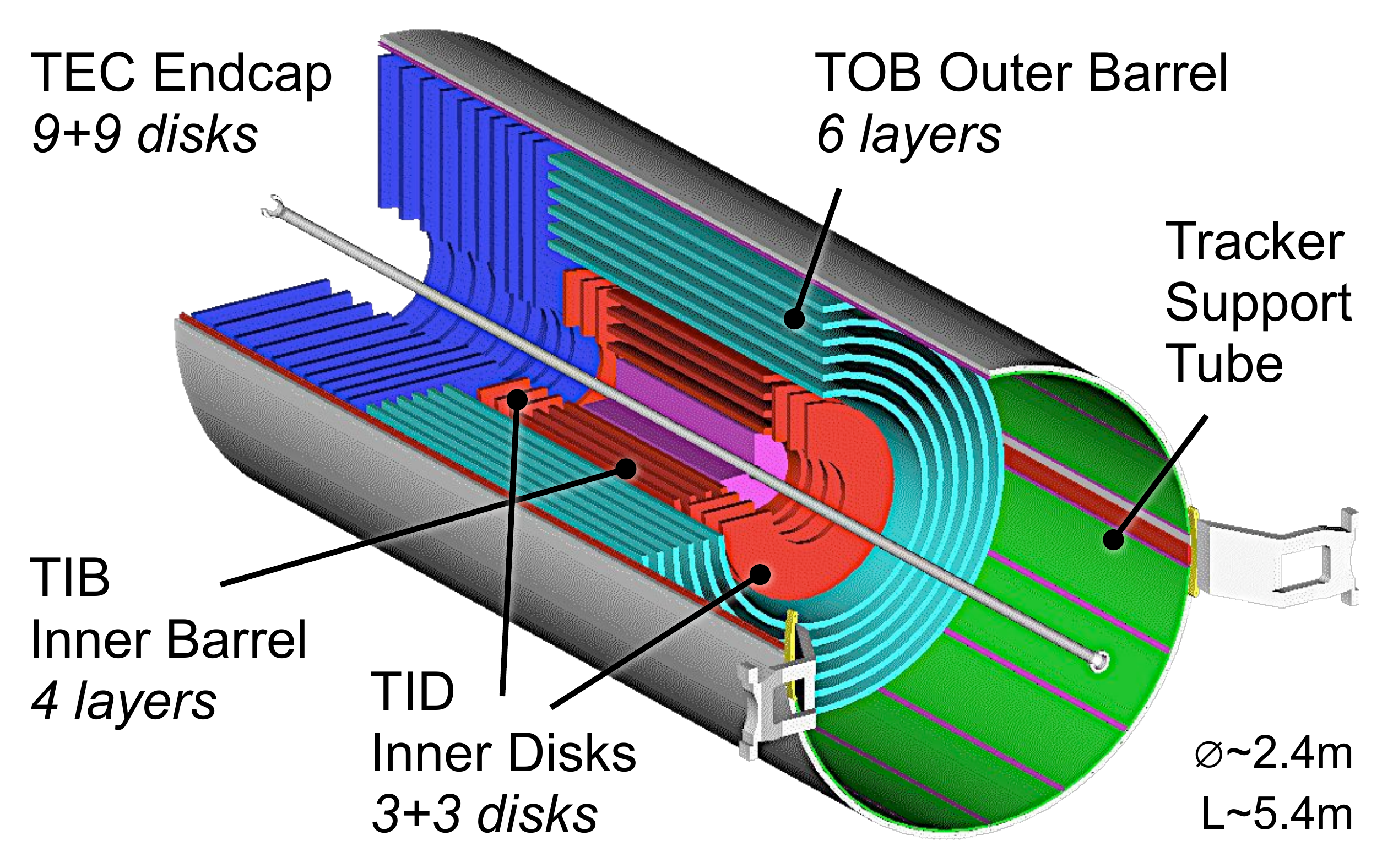}
\hskip 6mm
\includegraphics*[width=0.46\textwidth]{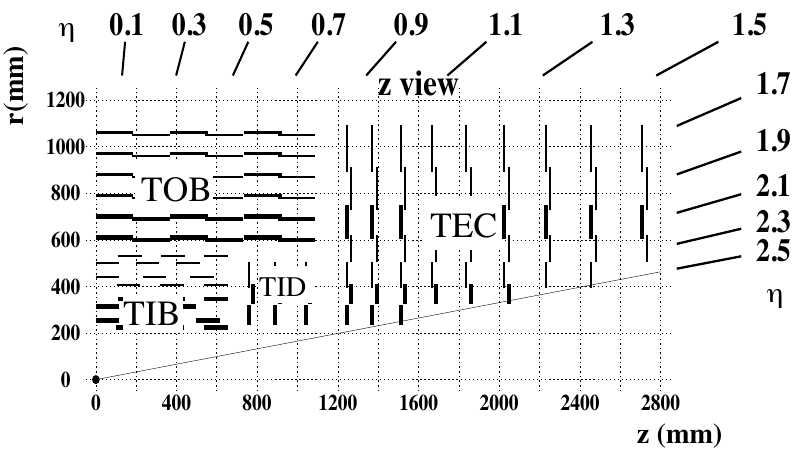}
\caption{The CMS SST: simplified view (left panel); a quadrant
of the $rz$ section (right panel, bold lines
represent double sided module assemblies).}
\label{fig:sst}
\end{center}
\end{figure}
The SST, shown in Figure~\ref{fig:sst},
covers the radial range between $20\cm$ and $110\cm$ around the LHC
interaction point. The barrel region ($| z |  < 110\cm$) is split into
a Tracker Inner Barrel (TIB) made of four detector layers, and a
Tracker Outer Barrel (TOB) made of six detector layers. The TIB is
complemented by three Tracker Inner Disks 
per side (TID). The forward and backward
regions ($120\cm < |z| < 280\cm$) are covered by nine Tracker
End-Cap (TEC) disks per side. In some of the layers and in the innermost
rings, special stereo modules (Fig.~\ref{fig:sst}, right panel) are
used to build double sided assemblies 
able to provide three-dimensional position measurement of the charged
particle hits. 

\begin{figure}[t]
\begin{center}
\includegraphics*[width=\textwidth]{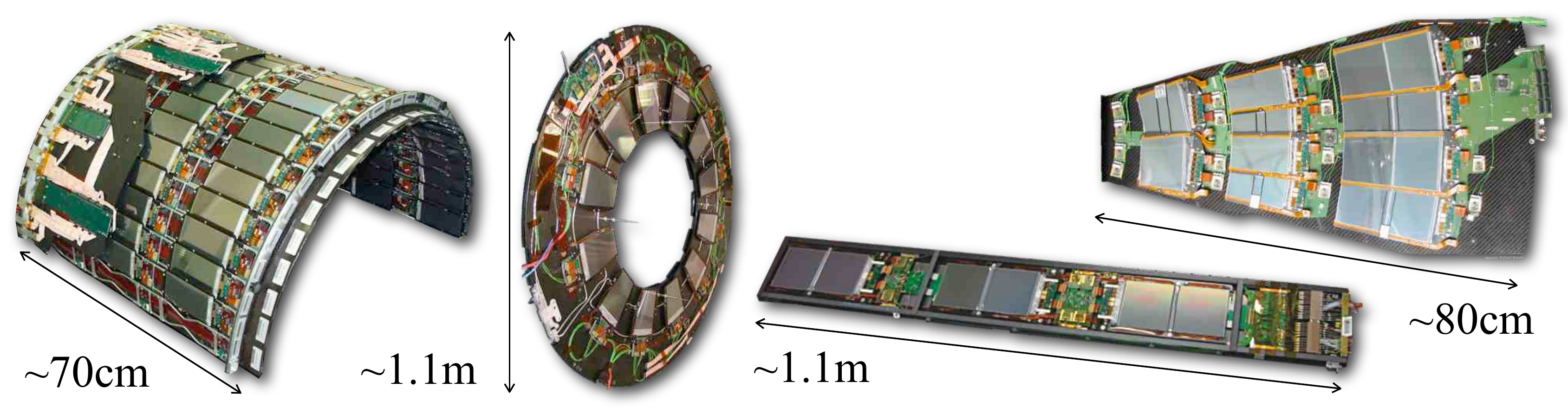}
\caption{From left to right: a shell of the TIB; the innermost
  ring of a TID disk; a rod of the TOB; a petal of the TEC.}
\label{fig:shell}
\label{fig:ring}
\label{fig:rod}
\label{fig:petal}
\end{center}
\end{figure}

The SST is made up of relatively small carbon fiber assemblies supporting
several modules to ease the handling and the mounting. These
substructures are fully equipped with cooling and 
readout electronics to allow for stand-alone tests to be
performed. This modularity has been the key-factor for the successful
assembly of the huge number of Si-strip modules~\cite{modules} and all
other ancillary components into the final detector.  
The following subdetector-dependent substructures can be found in the
SST (Fig.~\ref{fig:shell}): the TIB is split into 16 half cylinder {\em shells}
each holding 135 to 216 modules; the TID is
structured in $3$ {\em rings} per disk, each
designed to support 40 or 48 modules; the TOB is made of 688 {\em
  rods}, drawer-like structures providing support for 6
or 12 modules; and the TECs are made up of $144$ {\em petals} per endcap,
each holding 17 to 28 modules.  
During each stage of the assembly, every single component is verified. Optical
links are checked by measuring the output level of an auxiliary signal
issued by the readout chips and modules are checked using pedestal and
noise data at full bias ($400\V$) to spot high voltage issues and bad
channels~\cite{cc,smersi}. 
Once the substructures were completed, system-wide tests were
performed for which cooling was provided as in final CMS operating
conditions. Temperature 
probes mounted on the modules are effective for identifying cooling circuit
problems~\cite{palla}. In many cases, a cosmic ray setup was
implemented to measure the S/N ratio for MIPs and to perform track
reconstruction exercises, so to verify the overall mechanical precision
at the sub-millimeter level.

During 2006 all substructures were assembled to create the main
subsystems in regional integration centers. Finally the tracker
subsystems were installed in the Tracker Support Tube at the CERN
Tracker Integration Facility (TIF) between fall '06 and March '07. 
The tracker quality was excellent. The total
fraction of bad channels was found to be 0.21\%: 0.07\% due to module failures,
0.05\% due to bad optical links and 0.09\% of bad isolated channels, mostly
preexisting problems. The assembly procedures
and the stringent quality control tests have been proved to be sound
as the integration process caused almost no new
defects.

\begin{figure}[b]
\begin{center}
\includegraphics*[width=\textwidth]{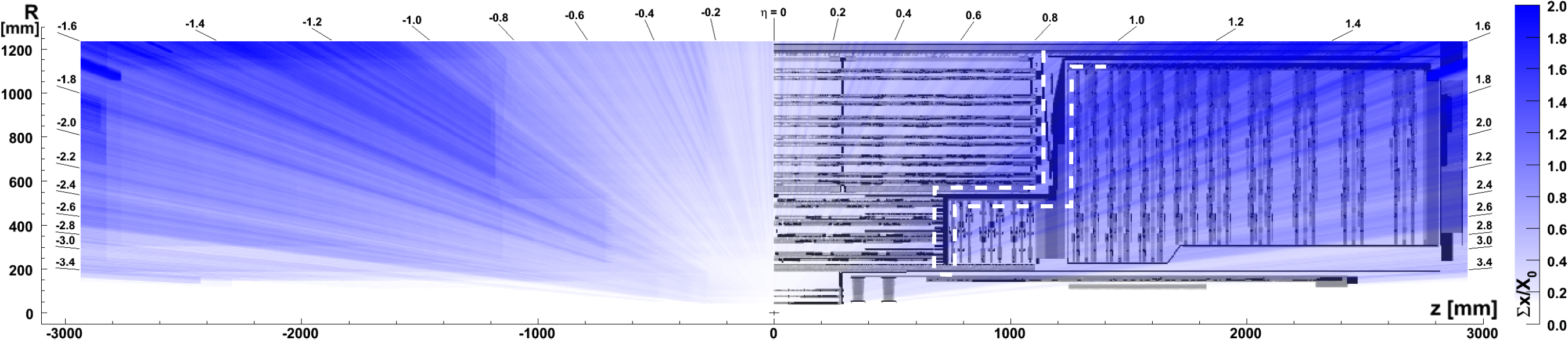}
\caption{Material budget $rz$ map as seen by straight tracks 
  originating at the interaction point: blue tone represents
  average of total upstream $x/X_0$ (i.e. local density of photon
  conversion tracks); gray tones (shown only on the $z>0$ part)
represents average of $1/X_0$ (i.e. local normalized
density of photon conversion vertexes). The white dashed line
indicates the region where TIB/TID services are deployed.} 
\label{fig:mb}
\end{center}
\end{figure}

Once the assembly of the SST was complete, it was possible
to evaluate the material budget by introducing into the
detector simulation all the last-minute details added during the
construction. The total thickness as a
function of $\eta$ reaches a maximum of about 1.8$X_0$ around
$1.2\lesssim|\eta|\lesssim1.8$ (Fig.~\ref{fig:mb}). This is because of
the presence of TIB and TID services in this region, which contribute
significantly to the overall material budget.
TIB and TID are the most challenging subsystems in terms of volumetric
density of channels (in units of $10^{6}$ channels per m$^3$: TIB~2.2, TID~1.1,
TOB~0.52, TEC~0.35) yielding a very large number of service  
connections in a very small space. This resulted in the previously
discussed material budget issue and in
difficulties to integrate the services in the limited room~allocated.

In December '07 the SST was installed underground in CMS and the full cabling and piping
was completed in March '08. Unfortunately two serious hardware failures
of the cooling system serving the tracker (November '07 and May '08)
prevented the checkout activities from being performed as scheduled,
i.e., immediately after the installation. The entire cooling system
needed to be refurbished and was available in June '08. 
In the 2008 LHC run, the SST will be conservatively operated at about $+12^\circ$C
instead of at the design temperature of $-10^\circ$C; in any case, the
latter is only required at design luminosity. During the 2008/2009 LHC
shutdown the cooling system will be commissioned for cold operations. 

\section{COMMISSIONING}

Immediately after the finalization of the assembly at the TIF,
an extensive testing program know as the ``Slice Test'' was carried out. It
consisted of cooling, powering and reading out approximately 15\% of  
the entire SST. A scintillator-based trigger permitted 5 million cosmic muon events 
to be collected at a rate of about 5Hz, filtering out low momentum muons by
means of a $5\cm$-thick lead shield placed on the 
bottom scintillator. The Slice Test allowed
the first experience to be gained on commissioning, safety, controls,
monitoring, data management, track reconstruction and alignment~\cite{tif}.

After the installation within CMS and once the cooling system was
available, the SST joined CMS operations. A cosmic global
run without magnetic field took place from 7 to 14 July '08. The test
was very important for CMS: the SST readout makes up approximately
$70\%$ of the entire 
DAQ system. Despite the tight schedule for the first round of
commissioning, the SST
performance is remarkable. The SST was smoothly read out
within the global CMS DAQ and several million events were
collected. $79\%$ of the SST modules were 
switched on (technical issues prevents the endcap on the $-z$ side
from being used) and, of those, around 93\% passed the testing
procedures first time. The vast majority of the small number of
modules excluded from this first global run will be simple to recover. 
In fact, the following subsequent more detailed commissioning
performed before the closure of CMS in fall '08, certified that the
fraction of bad modules in the entire SST is below $1\%$.  

\begin{figure}[t]
\begin{center}
\includegraphics*[width=0.4\textwidth]{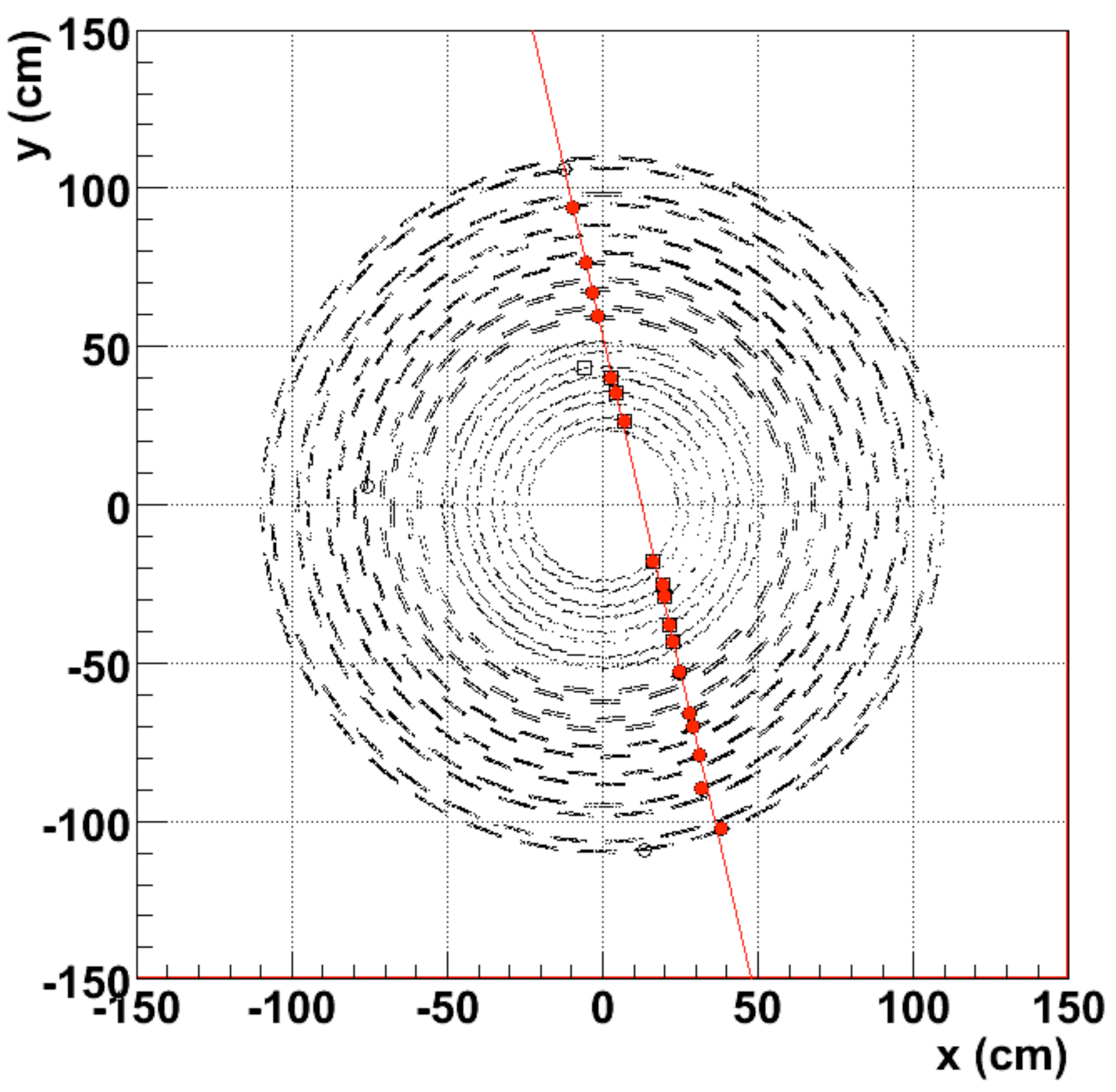}
\caption{A reconstructed cosmic muon track passing through the SST
  collected during the Global Cosmic Run.} 
\label{fig:hit}
\label{fig:track}
\end{center}
\end{figure}

Cosmic muon tracks have been observed in the global run: an example is
shown in Figure~\ref{fig:track}.

\section{CALIBRATION HIGHLIGHTS}

The optimization of the SST depends on several
calibration steps~\cite{tif}. The most relevant are briefly
described here.

\begin{figure}[t]
\begin{center}
\includegraphics*[width=0.4\textwidth]{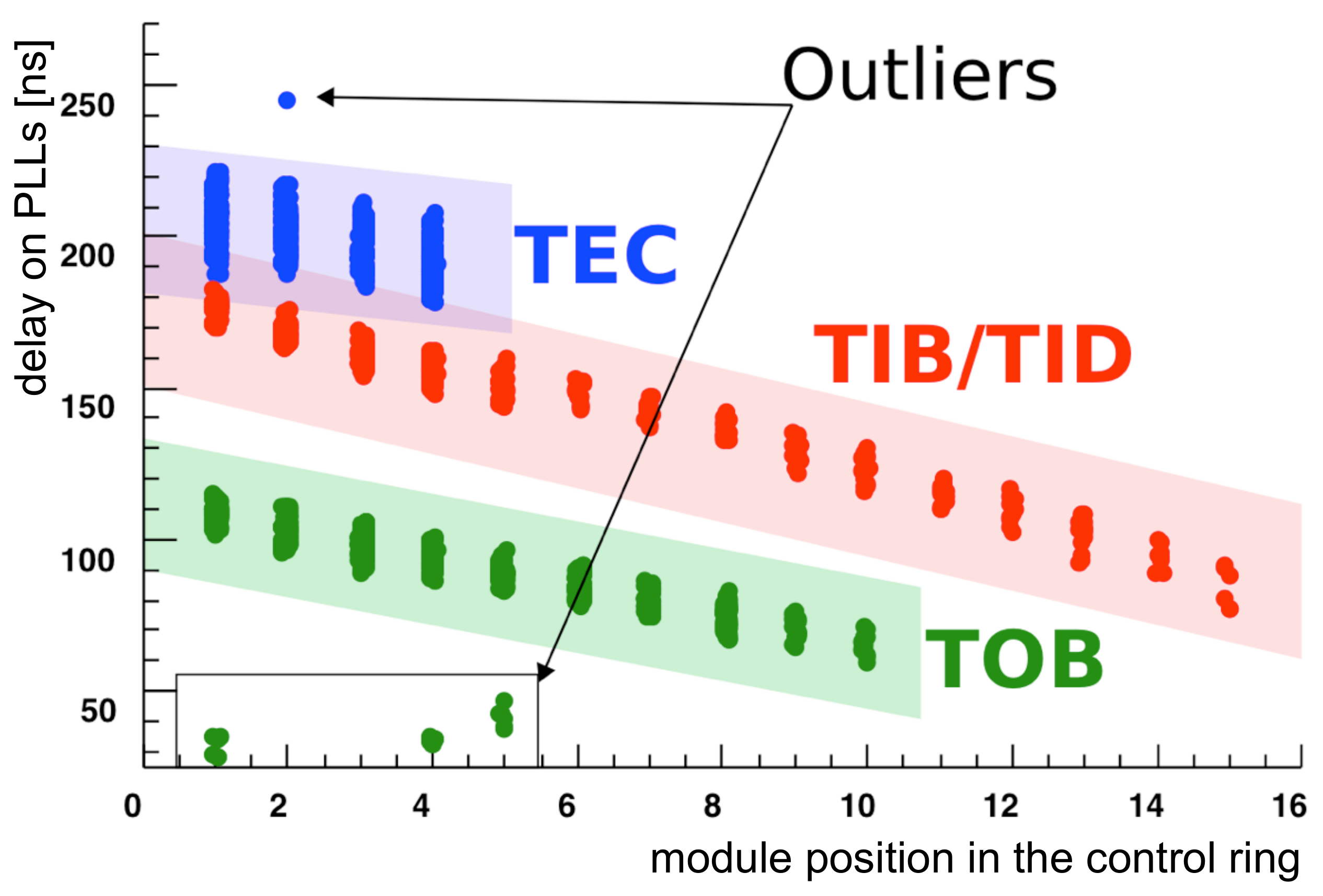}
\hskip 4mm
\includegraphics*[width=0.4\textwidth]{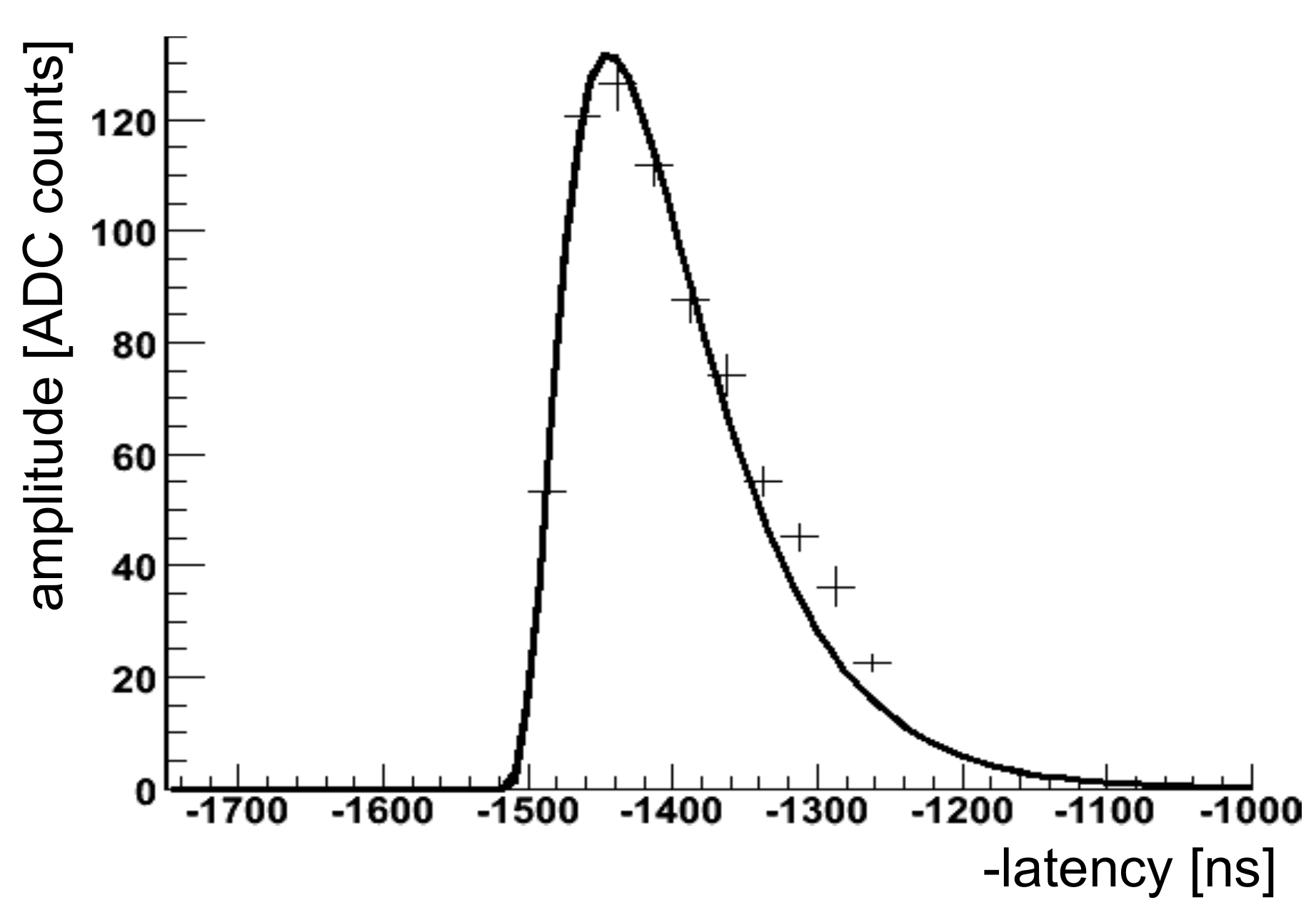}
\caption{The delays set on the modules programmable delay units against the
module position within the control ring (left panel); CR-RC pulse
shape reconstructed by scanning on the latency values (right panel).}
\label{fig:ta}
\label{fig:latency}
\end{center}
\end{figure}

The SST dimensions and cable/fiber lengths imply non negligible
timing differences compared to the LHC bunch-crossing period and the detector shaping time. 
Each module needs to be synchronized to compensate for path differences 
of control signals (and for any electronics delay) as shown in
Figure~\ref{fig:ta}, left panel, where the delays set on the
programmable delay unit of each module is plotted against
the number of upstream controller chips. In fact, to limit the
number of optical links, the control circuit
is implemented by daisy-chaining several controller chips 
introducing a delay that needs to be compensated for.

The analogue signal optimization is another fundamental calibration
step, to be obtained by setting up the working point 
parameters of
the front-end chips and the downstream optical chain. Among these, one
of the most important is 
the configuration of the front-end chips with the appropriate {\em latency}
value that, once a L1 trigger is issued, makes each module send the
data samples corresponding to the correct bunch crossing to the FEDs
(the SST ADC boards).
The result of the corresponding procedure ({\em latency scan}) are shown in
Figure~\ref{fig:latency}, right panel: the CR-RC pulse shape is reconstructed by
scanning a range of possible latency values in steps of $25\ns$ and
taking as the
optimal setting the closest value to the maximum
amplitude. In addition a fine tuning
within a single bunch crossing will compensate for time-of-flight
effects~\cite{tof}.

\section{CONCLUSIONS} 

The SST has been built, installed and read out within CMS.
Commissioning procedures, online and offline tasks have been 
demonstrated to work well. 
Final commissioning and checkout procedures have demonstrated that the SST quality is
excellent with more than $99\%$ of good modules. The SST is ready to
deliver optimally reconstructed~tracks.

\end{document}